\documentclass[%
 reprint,
superscriptaddress,
 amsmath,amssymb,
 aps,
]{revtex4-2}

\usepackage{graphicx}
\usepackage{dcolumn}
\usepackage{bm}
\usepackage{color}
\usepackage{gensymb}

\usepackage[normalem]{ulem}

\newcommand{\araa}{Annu. Rev. Astron. Astrophys.}
\newcommand{\aap}{A\&A}
\newcommand{\aj}{AJ}

\newcommand{\apjl}{ApJ Lett.}
\newcommand{\cqg}{CQG}
\newcommand{\planss}{Planet. Space Sci.}

\newcommand{\mnras}{MNRAS}

\newcommand{\Msun}{M$_\odot$}

\definecolor{amber}{rgb}{1.0, 0.75, 0.0}

\definecolor{amethyst}{rgb}{0.6, 0.4, 0.8}

\definecolor{blue-violet}{rgb}{0.54, 0.17, 0.89}

\definecolor{blue(munsell)}{rgb}{0.0, 0.5, 0.69}

\definecolor{byzantine}{rgb}{0.74, 0.2, 0.64}

\definecolor{carminered}{rgb}{1.0, 0.0, 0.22}

\definecolor{bittersweet}{rgb}{1.0, 0.44, 0.37}

\definecolor{ao(english)}{rgb}{0.0, 0.5, 0.0}

\begin{document}

\title{Hierarchical Triples as Early Sources of $r$-process Elements}

\author{I. Bartos}
\email{imrebartos@ufl.edu}
\affiliation{Department of Physics, University of Florida, Gainesville, FL 32611-8440, USA}
\author{S. Rosswog}
\email{stephan.rosswog@uni-hamburg.de}
\affiliation{Universit\"{a}t Hamburg, D-22761 Hamburg, Germany}
\affiliation{The Oskar Klein Centre, Department of Astronomy, Stockholm University, AlbaNova, SE-106 91 Stockholm, Sweden}
\author{V. Gayathri}%
\affiliation{Leonard E. Parker Center for Gravitation, Cosmology, and Astrophysics, University of Wisconsin–Milwaukee, Milwaukee, WI 53201, USA}
\affiliation{Department of Physics, University of Florida, Gainesville, FL 32611-8440, USA}
\author{M.C. Miller}
\affiliation{Department of Astronomy and Joint Space-Science Institute, University of Maryland, College Park, MD 20742-2421, USA}
\author{D. Veske}
\affiliation{Institut f\"{u}r Theoretische Physik, Philosophenweg 16, 69120 Heidelberg, Germany}
\author{S. Marka}
\affiliation{Department of Physics, Columbia University in the City of New York, New York, NY 10027, USA}

\begin{abstract}
Neutron star mergers have been proposed as the main source of heavy $r$-process nucleosynthesis in the Universe. However, the mergers' significant expected delay after binary formation is in tension with observed very early $r$-process enrichment, e.g., in the dwarf galaxy Reticulum II. The LIGO and Virgo gravitational-wave observatories discovered two binary mergers with lighter companion masses ($\sim 2.6$\,M$_\odot$) similar to the total mass of many binary neutron star systems in the Galaxy. The progenitor of such mergers could be a neutron star binary orbiting a black hole. Here we show that a significant fraction of neutron star binaries in hierarchical triples merge rapidly ($\gtrsim3\%$ within $\lesssim10$\,Myr after neutron star formation) and could explain the observed very early $r$-process enrichment. The neutron star binary can become eccentric via von Zeipel-Kozai-Lidov oscillations, promoting a fast coalescence followed later by a merger of the low-mass black hole with the higher-mass black hole in the system. We show that this scenario is also consistent with an overall binary neutron star merger rate density of $\sim100$\,Gpc$^{-3}$yr$^{-1}$ in such triples. Using hydrodynamic simulations we show that highly eccentric neutron star mergers dynamically eject several times more mass than standard mergers, with exceptionally bright kilonovae with an ``early blue bump" as unique observational signatures.
\end{abstract}

\maketitle

\section{Introduction}
\label{sec:introduction}

Based on observations in the Galaxy, there may be a dearth of compact objects in the so-called \textit{lower mass gap} of around $2.2-5\,M_\odot$ (\cite{1998ApJ...499..367B,2010ApJ...725.1918O,2011ApJ...741..103F}, but see \cite{2020A&A...636A..20W} for some contrary evidence).  Yet, the LIGO \cite{2015CQGra..32g4001L} and Virgo \cite{2015CQGra..32b4001A} observatories have discovered two such objects out of $\sim 100$ compact binary coalescences \cite{LIGOScientific2021djp}. In the gravitational wave event GW190814 \cite{2020ApJ...896L..44A} a compact object with mass $m_2=2.59^{+0.08}_{-0.09}$M$_\odot$ merged with a black hole of mass $m_1=23.2^{+1.1}_{-1.0}$M$_\odot$, while in the event GW200210\_092254 \cite{LIGOScientific2021djp} the merging masses were $m_2=2.83^{+0.47}_{-0.42}$M$_\odot$ and $m_1=24.1^{+7.5}_{-4.6}$M$_\odot$. The lower mass objects are likely black holes, albeit the possibility of them being neutron stars cannot be firmly excluded \cite{rhoades74,kalogera96}.

If isolated or binary evolution would normally {\it not} produce a lower mass gap (see \cite{2021ApJ...908L..38A} for other evolutionary possibilities), then a compact object can still reach a mass within the gap by two means. First, a neutron star, formed with mass $\lesssim 2.2$\,M$_\odot$, grows through gas accretion from its environment. This can occur in a low-mass X-ray binary or due to bound supernova ejecta \cite{2020ApJ...899L..15S}.  Another possibility is accretion from an Active Galactic Nucleus (AGN) disk, after which the compact {\bf object} could later merge with a black hole that is also in the disk \cite{2020ApJ...901L..34Y}. Second, a neutron star can merge with an object such as another neutron star. This can occur, followed by a secondary merger with a black hole, through assembly in a dense stellar cluster \cite{2021MNRAS.502.2049L}, in AGN disks \cite{2020ApJ...901L..34Y}, and in isolated stellar triples \cite{2021MNRAS.500.1817L} and quadruples \cite{2021MNRAS.502.2049L}.

The accretion and merger scenarios result in different mass distributions within the mass gap. Notably, based on the narrow mass distribution of known Galactic binary neutron star systems \cite{1994PhRvL..73.1878F,2012ApJ...757...55O,2013ApJ...778...66K,2019ApJ...876...18F} ($\approx 2.65\pm0.12$, \cite{2019ApJ...876...18F}), we can expect neutron star merger remnants to have masses around $\sim2.6$\,M$_\odot$, similar to the observed objects \cite{2021MNRAS.500.1817L}. Accretion, on the other hand, is likely to result in a more spread out mass distribution within the gap.

\begin{figure}[b]
    \centering
   \includegraphics[width=0.47\textwidth,,trim={4.8cm 6cm 16.8cm 7.8cm},clip]{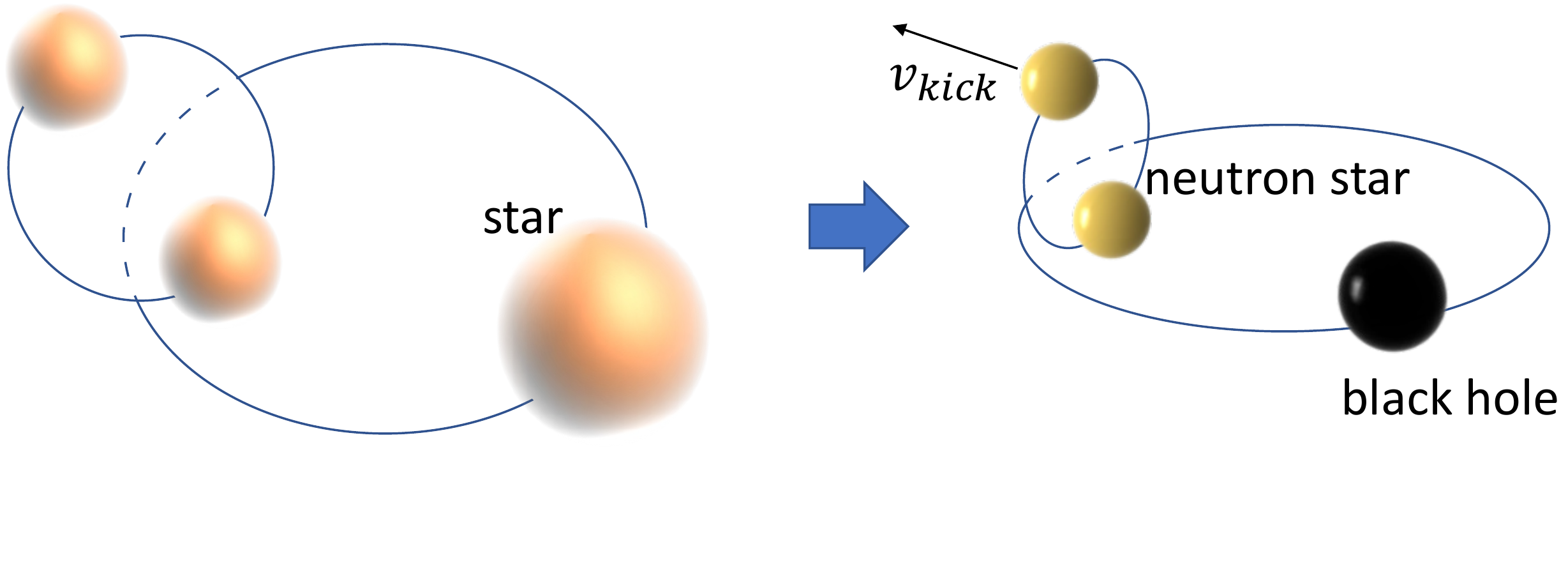}
    \vspace{-0.7cm}
    \caption{{\bf Illustration of the evolution of a hierarchical triple.} First, the most massive star on the outer orbit forms a black hole. Second, one of the less massive stars in the inner binary forms a neutron star. Finally, the third star, in the inner binary, forms a neutron star. The natal kick velocity $v_{\rm kick}$ the neutron star receives in this last step has the most significant impact on the triple's orbital evolution.} 
    \label{fig:illustration}
\end{figure}

Here we focus on the isolated stellar triple scenario, motivated by the masses of the objects observed in the mass gap that suggests a binary neutron star merger origin, and the difficulty of alternative scenarios accounting for the apparent rate of the secondary merger with a black hole (e.g., \cite{2021MNRAS.500.1817L}). 

We first present our prescription for the evolution of the triple system (Section \ref{sec:evolution}), followed by our results on the rapid merger  of the inner binary neutron star system (Section \ref{sec:rapidBNS}) and the rate of secondary mergers (Section \ref{sec:secondary}). We then focus on eccentric neutron star mergers that could by produced by triple systems, in particular the observational consequences of their unique mass ejection (Section \ref{sec:massejection}). We conclude in Section \ref{sec:conclusion}.

\section{Evolution model of the triple system} \label{sec:evolution}

The triple system initially contains three stars (see Fig. \ref{fig:illustration}), with two lighter stars forming the inner orbit with semi-major axis $a_1$, and with a more massive star on an outer orbit with semi-major axis $a_2$. We assume that the initial orbits are coplanar and circular. The more massive star evolves the most rapidly, eventually leaving behind a black hole with mass $m_2$. During formation, this black hole can receive a small kick velocity that makes the outer orbit somewhat eccentric. We neglected this initial kick for simplicity, noting that its expected small amplitude probably has limited effect, and its effect probably only enhances the merger rate we discuss below. In addition, the outer orbit can circularize due to tidal effects in the still remaining stellar binary. 

Subsequently, the two less massive stars undergo supernova explosions and leave behind two neutron stars. The neutron stars receive natal kicks at birth (e.g., \cite{2021MNRAS.500.1817L}). Here the kick of the first neutron star plays a much smaller role in the final evolution of the triple system as it occurs when its companion in the inner binary is still a star with an order of magnitude more mass than a neutron star. We therefore only consider the effect of the second neutron star's natal kick, $v_{\rm kick}$, on the triple system's evolution. We assumed that this natal kick's direction is isotropically distributed \cite{2018ApJ...855...82B}. 

The speed distribution of natal kicks is not well known, and both high velocities of a few hundred km\,s$^{-1}$ \cite{2005MNRAS.360..974H,2010ApJ...721.1689W} low velocities of a few tens of km\,s$^{-1}$ occur \cite{2016MNRAS.456.4089B}.

The natal kick of the second neutron star will result in both inner and outer orbits becoming eccentric with eccentricities $e_1$ and $e_2$, respectively, while the inner orbit will have nonzero inclination $i$ compared to the outer orbit. 

\section{Rapid neutron star mergers} \label{sec:rapidBNS}

In hierarchical triple systems, the inner orbit can become highly eccentric due to von Zeipel-Kozai-Lidov (ZKL) oscillations \cite{1910AN....183..345V,1962P&SS....9..719L,1962AJ.....67R.579K}. This leads to rapid merger compared to the merger timescale for a circular or low-eccentricity binary, which could account for $r$-process nucleosynthesis found in the early Universe \cite{2018PASA...35...17B}. In our model ZKL oscillations do not affect the initial triple systems since we assume that the initial orbits are co-planar. 

Without any eccentricity, a binary neutron star with a semi-major axis of 0.1\,au merges in $\sim 6\times10^{12}$\,yr. The merger times of orbits whose eccentricities periodically change via ZKL oscillations are shortened by a factor $(1-e_{\rm max}^2)^3$ for high $e_{\rm max}$, which is the maximum eccentricity reached during oscillations \citep{Liu_2018}. We simulated ZKL oscillations to quadrupolar order \citep{2016ARA&A..54..441N} (the octupolar term is zero in our case due to the equal mass binary system). General-relativistic corrections to the eccentricity evolution will not be significant in our case as they will only affect a negligibly small fraction of the highest eccentricity systems \cite{2002ApJ...576..894M}. We included the effect of gravitational-wave emission in the evolution \citep{1964PhRv..136.1224P}. 

As a fiducial model we adopted an initial inner orbit semi-major axis $a_1=0.1$\,au and initial outer orbit semi-major axis of $a_2=1$\,au. We carried out Monte Carlo simulations of the binary evolution. In each realization we randomly selected the direction of $v_{\rm kick}$ from an isotropic distribution, the phase of the binary neutron star system compared to the outer binary within the same plane from a uniform distribution, and the magnitude of  $v_{\rm kick}$ from a uniform distribution between $0-400$\,km\,s$^{-1}$. For this test, for simplicity, we also assumed that $e_2=0$. We then carried out simulations to determine $e_{\rm max}$. 

We found that $\sim3\%$ of neutron star mergers occur within $10^7$\,yr following the formation of the second neutron star. This fraction can be even higher for some parameter choices. For example, the fraction is $\sim10\%$ if we assume a single velocity of $v_{\rm kick}=200$\,km\,s$^{-1}$.

\section{Rate of secondary mergers} \label{sec:secondary}

We used our triple evolution model to evaluate the relative rate of the primary neutron star mergers in triples to the rate of secondary mergers of a black hole with the remnant of the neutron star mergers. We carried out a Monte Carlo simulation with the same setup as discussed in the previous Section. 

In a given realization, we counted the inner binary as merged if two neutron stars remained gravitationally bound after the natal kick, and if they merged within $t_{\rm max}=10^{10}$\,yr after formation. The merger time was determined by accounting for gravitational wave emission, and, if the binary remained gravitationally bound to the outer black hole after the natal kick, for ZKL oscillations. For the latter case, we only considered realizations in which the inner binary remained bound to the black hole if $a_1< a_2/3$ after the natal kick, as for closer separations the triple system becomes chaotic.

We counted a realization to undergo secondary mergers if both the binary neutron star system and the outer binary remained gravitationally bound, and if both inner and outer binaries merged within $t_{\rm max}=10^{10}$\,yr after formation. For the outer binary's evolution we accounted for gravitational wave emission only. Since the outer binary's angular momentum is much greater than that of the inner binary, ZKL effects on the outer binary are negligible.

For our fiducial model with $a_1=0.1$\,au and $a_2=1$\,au prior to the natal kick, and with $v_{\rm kick}$ taken from a uniform distribution within $0-400$\,km\,s$^{-1}$, we found that the rate of black hole--mass gap mergers is $f=3\%$ of that of neutron star mergers in triple systems. This means that approximately 30 times more neutron star mergers are expected from triple systems than the observed rate density of $\sim10$\,Gpc$^{-3}$\,yr$^{-1}$ of secondary mergers. The implied rate $\sim 300$\,Gpc$^{-3}$\,yr$^{-1}$ of all double neutron star mergers, if this is the only pathway, is comparable to the total reconstructed rate density $105.5^{+190.2}_{-83.8}$\,Gpc$^{-3}$yr$^{-1}$ of neutron star mergers \cite{2021arXiv211103634T}. In this scenario, about 0.1\% of triple systems will produce black hole--mass gap mergers within $10^{10}$\,yr. To characterize the parameter dependence of this result, we show $f$ for select initial parameters in Table \ref{tab:fs}.


\begin{table}
\centering
\begin{tabular}{|c|c|c|c|c|}
\hline
$a_1$ [au] & $a_2$ [au] & $v_{\rm kick,min}$ [km/s] & $v_{\rm kick,max}$ [km/s] & $f$ [\%] \\ 
\hline
0.1 & 1 & 0 & 400 & 3 \\
\hline
0.1 & 3 & 0 & 400 & 1.5 \\
\hline
0.1 & 10 & 0 & 400 & 0.3 \\
\hline
0.2 & 1 & 0 & 400 & 0.1 \\
\hline
0.1 & 1 & 300 & 300 & 40 \\
\hline
\end{tabular}
\caption{{\bf Ratio $f$ of black hole--mass gap mergers and neutron star mergers} for different initial triple system parameters. Kick velocity probability density is isotropic and uniform within [$v_{\rm kick,min},v_{\rm kick,max}$].}\label{tab:fs}
\end{table}

While the initial triple parameters are uncertain, our results show that a significant fraction, or even majority, of all neutron star mergers may occur in triple systems.

\section{Mass ejection in highly eccentric neutron star mergers} \label{sec:massejection}

The eccentricity distribution of binary neutron stars due to ZKL oscillations can extend to high eccentricities. In comparable scenarios, $\sim10\%$ of the binaries were estimated to merge in close to head-on collisions \cite{2014ApJ...781...45A,2014MNRAS.439.1079A}. 

\begin{figure*}[htbp] 
   \centerline{
   \includegraphics[width=17.9cm,trim={0cm 0cm 1.2cm 0cm},clip]{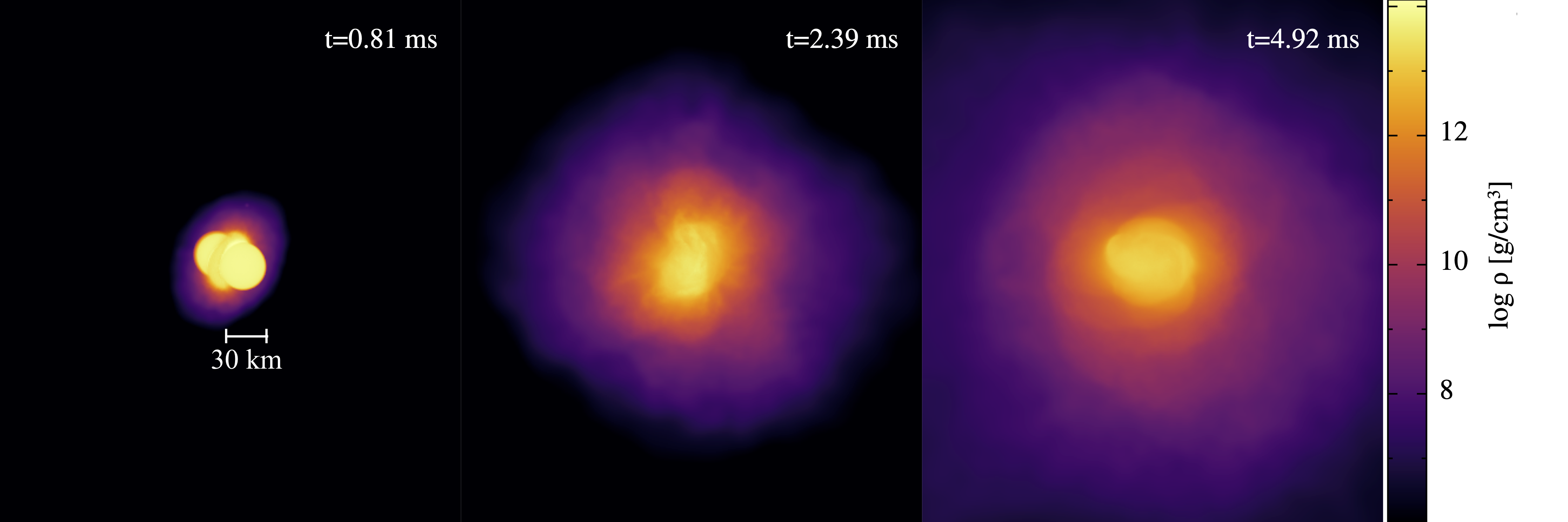}   }
   \caption{{\bf Volume rendering of a close to head-on ($\beta=5$) neutron star collision.} The neutron stars had masses of 1.3
   and 1.4\, M$_\odot$. Time (see legend) is measured from the simulation start (initial separation 100 km). Color indicates matter density. } 
   \label{fig:collision}
\end{figure*}

To investigate the multi-messenger observational consequences of such high-eccentricity collisions, we analyzed a set of numerical simulations presented in \cite{rosswog13b} to which we refer to for more technical details. The impact strength was defined as 
$\beta \equiv (R_1+R_2)/r_{\rm p}$, where the $R_i$ denote the 
stellar radii and $r_{\rm p}$ is the pericenter distance. These simulations focused on a quasi-circular merger with a 1.3 and a 1.4\, M$_\odot$ neutron star and contrasted it
with a "grazing" ($\beta=1$) and a close-to-central ($\beta=5$) collision of two neutron stars of the same masses. As an example, Fig.~\ref{fig:collision} shows a volume rendering of the $\beta=5$ collision. 

The simulations included a tabulated nuclear equation of state \cite{shen98b} and a multi-flavour neutrino leakage scheme \cite{rosswog03a}, but they were not fully general relativistic. Therefore, we expect the corresponding collisions in nature to be even more violent, and in particular to produce higher temperatures, higher electron fractions $Y_{\rm e}$
and therefore more lower opacity ejecta \cite{kasen13a,tanaka20a,rosswog22c}. The neutrino leakage scheme adopted in the simulations allows for an increase of
$Y_{\rm e}$ via positron captures, $e^+ + n  \rightarrow p +\bar{\nu}_e$, but it does not account for neutrino absorption. For these reasons we consider the $Y_{\rm e}$ obtained in these simulations as robust lower limits and therefore nature's electromagnetic transients from non-circular encounters are expected to have an even brighter early blue emission component than what we show here.

In order of increasing collision violence (merger $\rightarrow \beta=1 \rightarrow 
\beta=5)$, we found that the $Y_e$-distribution is systematically shifted towards larger values, and, for the reasons discussed above, we consider this trend as very robust and even stronger in nature (see Table~\ref{tab:ejecta}).
When analyzing the {\em dynamic} ejecta, we split them into two groups according 
to whether they are below (``low-$Y_e$") or above (``high-$Y_e$ component")
the critical electron fraction value 
$Y_e^{\rm crit}=0.25$ \cite{korobkin12a,lippuner15,rosswog18}, where the composition
(and therefore the resulting opacity) changes abruptly.

Secular ejecta from accretion disks are expected to dominate the total ejecta mass and the
simulations by \cite{rosswog13b} show that the accretion disk masses are comparable for the different considered encounters (see Tab.~\ref{tab:ejecta}).  We did not separately simulate the secular ejection process, but instead assumed that 35\% of the accretion disk mass will be ejected \cite{beloborodov08,metzger08a,siegel17a,siegel18,fernandez19,miller19}
and will expand at $0.1\,c$. Since the secular ejecta's electron fraction is currently an unsettled question, we explored the cases where they are above (0.3) and below (0.2) the critical electron fraction value $Y_e^{\rm crit}$. The obtained ejecta properties are given in Tab.~\ref{tab:ejecta}. The dynamical ejecta of the collisions exceed those of the circular mergers by factors of a few, while the disk masses/secular ejecta are of the same order.
\begin{table*}[]
    \centering
    \begin{tabular}{|c|c|c|c|}
    \hline
      encounter & dynamic, low-$Y_e$ &  dynamic, high-$Y_e$ & secular\\
                & $m$, $\langle v_\infty \rangle$, $\langle Y_e \rangle$  
                & $m$, $\langle v_\infty \rangle$, $\langle Y_e \rangle$
                & $m$, $\langle v_\infty \rangle$, $\langle Y_e \rangle$\\
                \hline
         merger & 0.0138, 0.117, 0.047 & 1.42E-4, 0.20, 0.275 & 0.088, 0.1, 0.2 \& 0.3 \\      \hline
         collision $\beta=1$ & 0.0580, 0.125, 0.049 & 1.92E-3, 0.237, 0.303 &  0.095, 0.1, 0.2 \& 0.3 \\         \hline
         collision $\beta=5$ & 0.0198, 0.25, 0.18 &   0.0102, 0.33, 0.32  & 0.112, 0.1, 0.2 \& 0.3 \\
         \hline
    \end{tabular}
    \caption{Masses of the different ejecta components. All encounters are between a
    1.3 and a 1.4 \Msun neutron star.}
    \label{tab:ejecta}
\end{table*}

\begin{figure*}
   \centerline{
   \includegraphics[width=8.5cm]{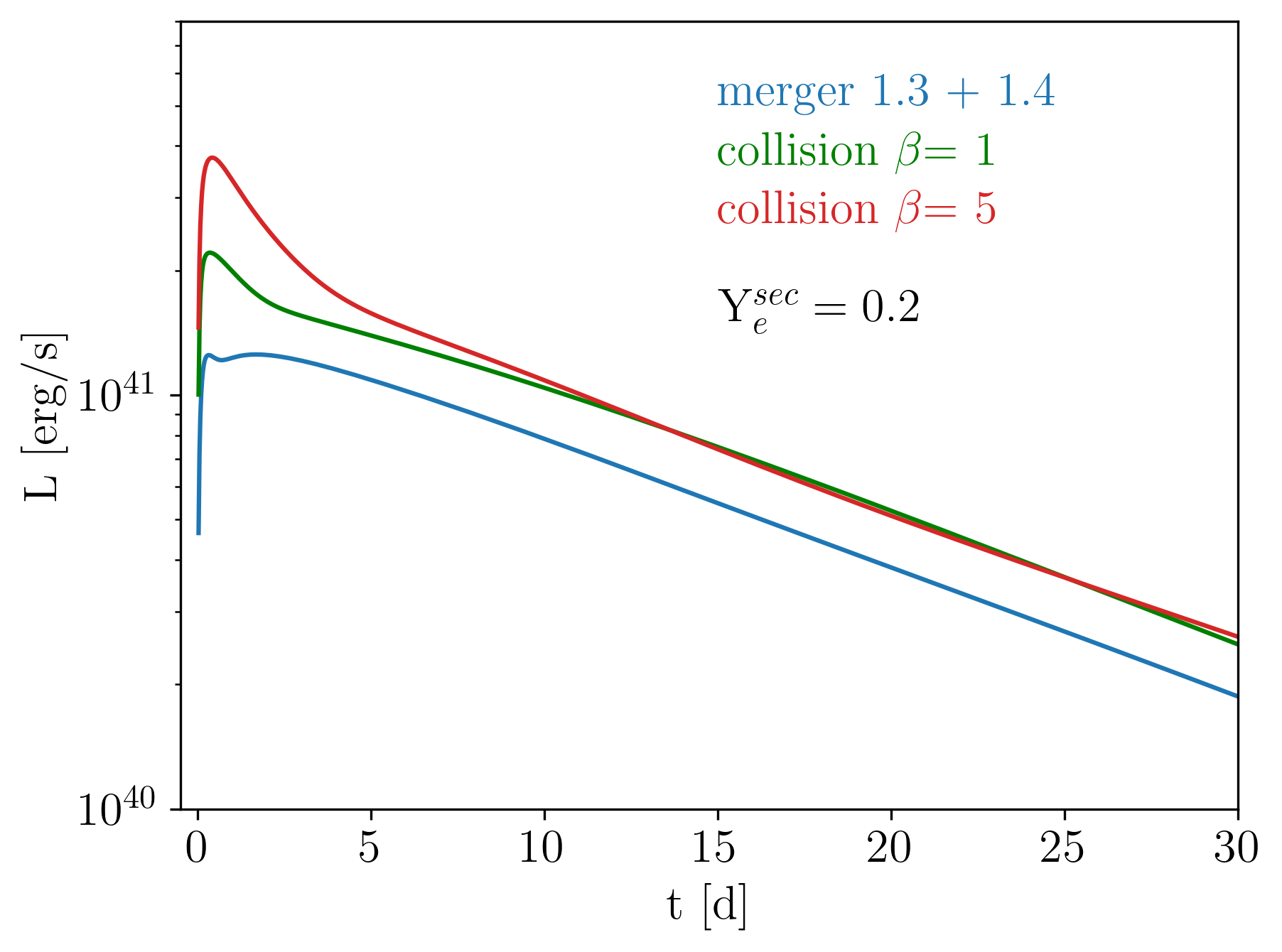} 
   \hspace*{-0.3cm}
   \includegraphics[width=8.5cm]{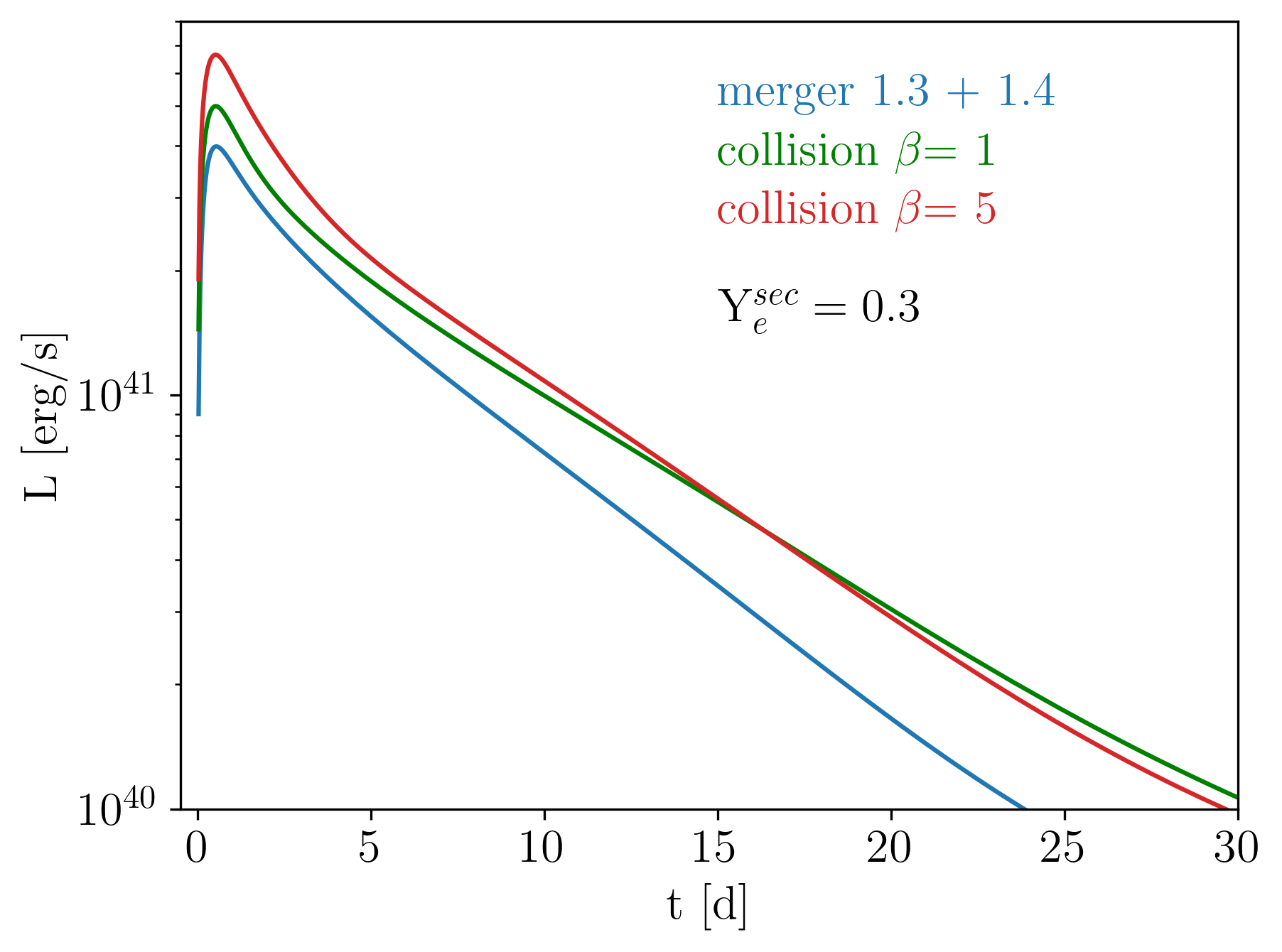}   }
    \vspace{-0.4cm}   
   \caption{{\bf Bolometric kilonova light curves}. We show light curves for quasi-circular as well as highly eccentric mergers (see legend). We separately show results for disk electron fractions $Y_e^{sec}=0.2$ (left) and $Y_e^{sec}=0.3$ (right).}
   \label{fig:bolometric_Ye0.2_Ye0.3}
\end{figure*}

Based on the properties of these three ejecta components, we computed the kilonova light 
curves with a semi-analytic eigenmode expansion model \cite{wollaeger18,rosswog18} which is based on \cite{pinto00}. Compared to \cite{rosswog13b}, here we selected the average opacity 
according to the electron fraction following \cite{tanaka20a} (their Table 1), we used nuclear
heating rates from the nuclear network calculations of \cite{rosswog22c} and we
applied {\bf a} thermal efficiency that is based on \cite{barnes16} with parameters suggested in
\cite{metzger19a}. 
We show the obtained light curves in Fig.~\ref{fig:bolometric_Ye0.2_Ye0.3}. We see that, in order of collision violence (merger $\rightarrow \beta=1 \rightarrow \beta=5$) the light curves become brighter and show a more pronounced "early blue bump".

\section{Conclusion} \label{sec:conclusion}

We found that the gravitational wave events GW190814 and GW200210\_092254 suggest the possibility of a high fraction of neutron star mergers that occur in hierarchical triples. A significant fraction (few \%) of these mergers could be highly eccentric and merge rapidly after neutron star formation, within $\lesssim 10$\,Myr. 

Neutron star mergers are the only observed source with $r$-process production \cite{LIGOScientific:2017vwq}, with a measured yield consistent with these mergers being the dominant $r$-process source in the Universe. However, neutron star mergers are expected to be delayed compared to binary formation as binary separation is typically reduced only via gravitational waves \cite{2014ARA&A..52...43B}. This delay is in tension with early $r$-process enrichment, observed in the dwarf galaxy Reticulum II \cite{2016Natur.531..610J} and in early stars in the Galaxy \cite{2018IJMPD..2742005H,2017ApJ...836..230C}, leading to suggestions that collapsars, which can occur shortly after star formation, may be important $r$-process sources \cite{2019Natur.569..241S}. 

Our results reconcile the observed early $r$-process enrichment with neutron star mergers as the main source of heavy elements by showing that a sizable fraction of mergers occur rapidly, effectively tracking star formation.

An observational prediction of the hierarchical triple scenario examined here is a significant rate density ($\sim 10$\,Gpc$^{-3}$yr$^{-1}$) of highly eccentric neutron star mergers that could be detectable with gravitational waves. These mergers also result in higher dynamical ejecta masses and higher $Y_{\rm e}$ than quasi-circular mergers, producing pronounced  ``early blue bumps" during the first days of the resulting kilonova light curve, which could be identifiable via optical follow-up observations.

\section{Acknowledgments}
It is a pleasure to acknowledge insightful discussions with Cecilia Chirenti and Friedrich-Karl Thielemann.
I.B. acknowledges the support of the University of Florida, the Alfred P. Sloan Foundation, and NSF grants PHY-1911796, PHY-2110060 and PHY-2207661. 
SR has been supported by the Swedish Research Council (VR) under 
grant number 2020-05044, by the research environment grant
``Gravitational Radiation and Electromagnetic Astrophysical
Transients'' (GREAT) funded by the Swedish Research Council (VR) 
under Dnr 2016-06012, by the Knut and Alice Wallenberg Foundation
under grant Dnr. KAW 2019.0112,   by the Deutsche 
Forschungsgemeinschaft (DFG, German Research Foundation) under 
Germany’s Excellence Strategy – EXC 2121 ``Quantum Universe'' 
– 390833306 and by the European Research Council (ERC) Advanced 
Grant INSPIRATION under the European Union’s Horizon 2020 research 
and innovation programme (Grant agreement No. 101053985).
SR's calculations were performed 
on the facilities of the North-German Supercomputing Alliance (HLRN), 
Sz.M. is grateful for the generous support of Columbia University. D.V. was supported by European Research Council (ERC) under the European Union’s Horizon 2020 research and innovation programme grant agreement No 801781.  
 G.V. acknowledges
the support of the National Science Foundation under grant PHY-2207728.

\bibliographystyle{apsrev4-1}

\bibliography{Refs}

\end{document}